\documentclass[namedreferences]{solarphysics}
\usepackage[optionalrh]{spr-sola-addons} 
\usepackage{graphicx}        
\usepackage{color}           
\usepackage{url}             
\usepackage{soul}

\newcommand{\kms}{~km~$s^{-1}$}

\newcommand{\atrous}{\emph{\`{a} trous}}
\newcommand{\app}{$\sim$}
\newcommand{\dns}{DN $s^{-1}$}
\newcommand{\arcsec}{"}

\newcommand{\aap}{    {\it Astron. Astrophys.}}

\newcommand{\apj}{    {\it Astrophys. J.}}

\newcommand{\solphys}{{\it Solar Phys.}}

\begin{document}

\begin{article}

\begin{opening}

\title{An improved method for estimating the velocity field of coronal propagating disturbances}

\author{Huw~\surname{Morgan}$^{1}$}
\author{Marianna~\surname{Korsos}$^{1}$}
\runningauthor{Morgan \& Korsos}
\runningtitle{A method for estimating coronal velocity fields}

   \institute{$^{1}$ Adran Ffiseg, Prifysgol Aberystwyth, Ceredigion, Cymru, SY23 3BZ\\
                     email: \url{hmorgan@aber.ac.uk} \\ 
             }

\begin{abstract}
The solar corona is host to a continuous flow of propagating disturbances (PD). These are continuous and ubiquitous across broad regions of the corona, including the quiet Sun. The aim of this paper is to present an improved, efficient method to create velocity vector field maps, based on the direction and magnitude of the PD as observed in time series of extreme ultraviolet (EUV) images. The method is presented here for use with the Atmospheric Imaging Assembly (AIA)/Solar Dynamics Observatory (SDO) EUV channels, and takes as input \app2 hours of images at the highest 12s cadence. Data from a region near disk center is extracted, and a process called time normalization applied to the co-aligned data. Following noise reduction using \atrous\ decomposition, the PD are effectively revealed. A modified Lucas Kanade algorithm is then used to map the velocity field. The method described here runs comfortably on a desktop computer in a few minutes, and offers an order of magnitude improvement in efficiency compared to a previous implementation. Applied to a region of the quiet Sun, we find that the velocity field describes a mosaic of cells of coherent outwardly-diverging PD flows, of typical size 50 to 100\arcsec\ (36 to 72Mm). The flows originate from points and narrow corridors in the cell centres, and end in the narrow boundaries between cells. Visual comparison with ultraviolet AIA images shows that the flow sources are correlated with the bright photospheric supergranular network boundaries.  Assuming that the PD follow the local magnetic field, the velocity flow field is a proxy for the plane-of-sky distribution of the coronal magnetic field, and therefore the maps offer a unique insight into the topology of the corona. These are particularly valuable for quiet Sun regions where the appearance of structures in EUV images is hard to interpret.
\end{abstract}
\end{opening}

\section{Introduction}

\cite{morgan2018} present a method called Time Normalised Optical Flow (TNOF), which revealed the presence of propagating disturbances (PD) across broad regions of the corona, including part of a small active region and quiet Sun. Their method was based on two main steps - a processing step where a time series of on-disk extreme ultraviolet (EUV) images was filtered and modified (time normalisation), and a characterisation step where a vector velocity field was estimated based on the PD motions (optical flow). Whilst effective, the method was complicated, cumbersome and computationally inefficient, taking several hours to run on a desktop computer. 

The PD described by \cite{morgan2018} are faint - with typical amplitudes of at most \app4 \dns, or less than 2\%\ of the background signal, and not visible in unprocessed time series. They are ubiquitous in the quiet Sun and active regions, and continuous in the sense that PDs appear quasi-periodically and propagate along a similar direction to past PDs in the same region. This characteristic is supported by previous studies: for example, \cite{stenborg2011} state that disturbances are a coronal phenomenon that exist permanently everywhere, and \cite{wang2009} found similar PD in a fan-like loop structure. The propagation velocities are on the order of tens of \kms, and have quasi-periodicities, or repetition times, of a few minutes. Most studies interpret the PD as slow magnetoacoustic waves, with several emphasising propagation along open magnetic field \citep{stenborg2011}. \cite{morgan2018} was able to reveal and characterise fainter events compared to other studies, and found PD in closed field structures in an active region and quiet Sun, with the strongest signatures in the active region loops.

This paper presents an improved, efficient method to mapping the velocity field of coronal PD, and shows that the resulting velocity vector fields give a unique mapping of the structure of the quiet corona. An overview of the method, with detail of the new, efficient, optical flow technique is given in section \ref{method}. Results are presented in section \ref{results}, with a discussion of the velocity vector field in the context of the quiet Sun magnetic topology. A summary is given in section \ref{summary}

\section{Method} 
\label{method}

\subsection{Observations and preprocessing}
\begin{figure}    
\centerline{\includegraphics[width=1.0\textwidth,clip=]{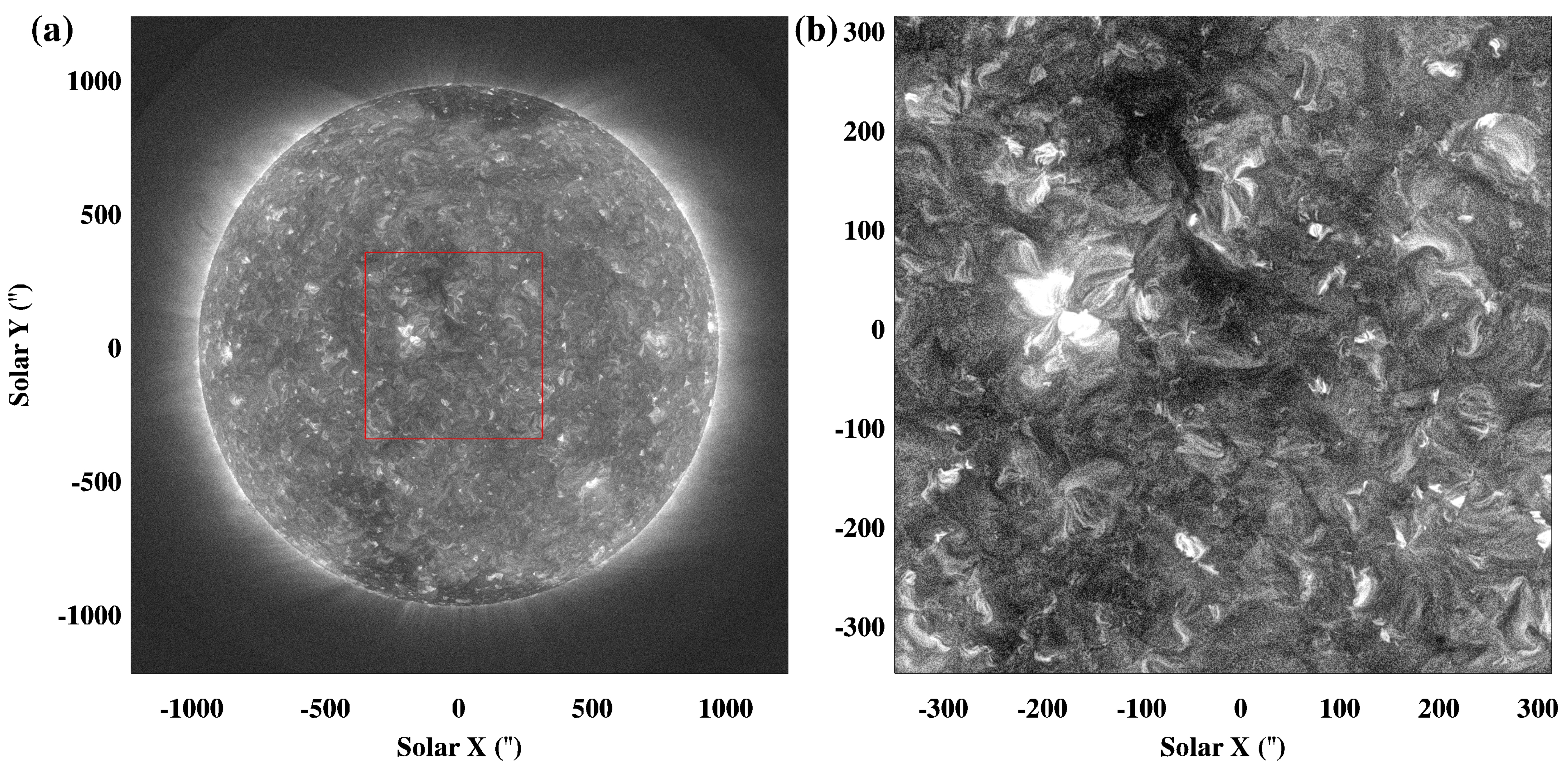}}
\caption{(a) Full AIA 193\AA\ image of 2019/11/06 14:00, with the red box bounding the region of study. (b) More detail of the region of study. Both these images have been processed using Multiscale Gaussian Normalisation to enhance detail \citep{morgan2014}.}
\label{fig1}
\end{figure}

The method is applied here to a time series of 193\AA\ EUV images taken by the Atmospheric Imaging Assembly (AIA, \opencite{lemen2011}) aboard the Solar Dynamics Observatory (SDO, \opencite{pesnell2012}) during 2019/11/06. The spatial pixel size of the data is 0.6\arcsec, and the cadence is regular at 12s. The method requires around 2 hours of observation (600 images) to give clean results, thus data spanning from 2019/11/06 13:00 to 15:00 are used here. The two hour span is a compromise between maximising the number of time steps available for the method to calculate a velocity, thus maximising the statistics; whilst maintaining a region near disk center without any large structural changes in the corona. Decreasing this two-hour span leads to velocity fields that are less coherent - thus there are less time steps over which to determine a dominant velocity, and in the resulting velocity fields we see increasingly large variation on small spatial scales and decreasing correlation with the underlying coronal structure. Figure \ref{fig1}a shows a full image taken by the AIA 193\AA\ channel for the midpoint of this time period. We restrict the application of the method to a region centered on disk center, as indicated by the red box in figure \ref{fig1}a. Figure \ref{fig1}b shows greater detail of this region. There are practical reasons to choose a region of interest near disk center: we can use the Solar Software cutout service to request and download data of manageable size, and computer memory puts constraints on the size of the region during the method process. Another reason is that the results are easier to interpret for disk center regions, since estimated plane-of-sky velocities are geometrically closer to being parallel to the solar surface. 

\begin{figure}    
\centerline{\includegraphics[width=1.0\textwidth,clip=]{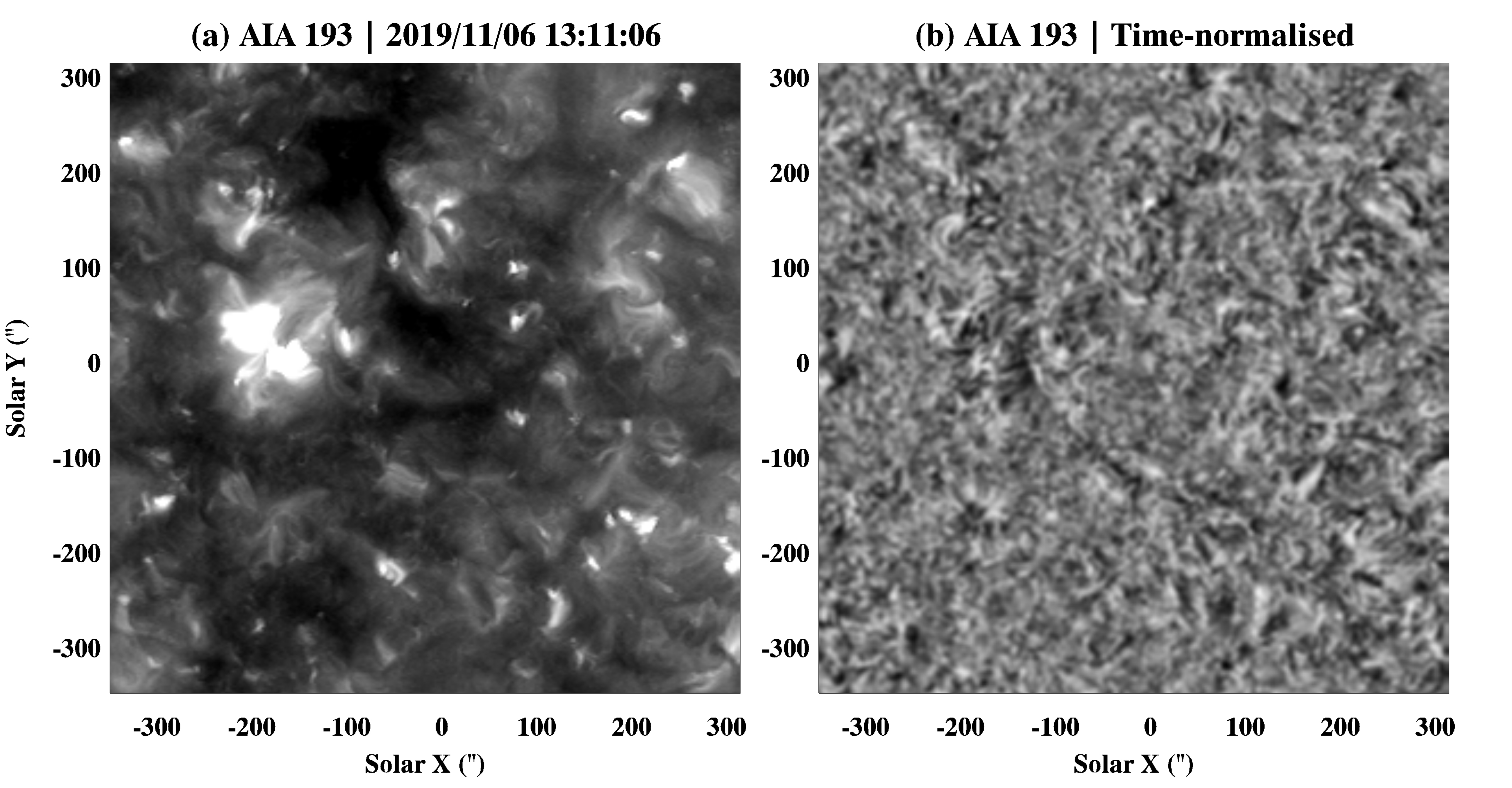}}
\caption{(a) The region of interest with intensities scaled by a simple square root (Gamma transform). (b) The corresponding time-normalised image. A movie accompanies this figure, showing the time series for both (a) unprocessed and (b) processed images.}
\label{fig2}
\end{figure}

The data, cropped to the region of study, is opened as a datacube of dimensions $[x,y,t]$, with each image's intensity divided by the exposure time to give units of \dns. The first step is to align the images over time. A Fourier transform correlation-based procedure calculates the global translational $[x,y]$ shifts of each image from the previous image to subpixel accuracy \citep{fisher2008}. The cumulative sum of these shifts over time are fitted to $2^{nd}$ degree polynomials. Each image is then shifted, using interpolation, by the appropriate $[x,y]$ shift so that all images are aligned to the image taken closest to the central time (in this case, 2019/11/06 14:00). The translational shifts lead to missing data at the left and right margins of the data - margins that increase in width towards the start and end of the time series. These margins are identified and discarded from the whole datacube so that each pixel has a complete time series. An image from the resulting aligned datacube is shown in figure \ref{fig2}a, and the accompanying movie shows the whole time series. Following alignment, the datacube is rebinned spatially so that the average value over each $4 \times 4 = 16$ input pixels form a new pixel (typically, the size of the input image region is $800 \times 800$ pixels, and the rebinned image becomes $200 \times 200$ pixels). Rebinning the data in this way is necessary to reduce noise and is computationally efficient. Tests on full-resolution data (made on a smaller spatial region due to computational limitations) lead to noisier velocity fields, with less coherence and large variations on smaller spatial scales.

\subsection{Time normalisation}
The next steps apply time normalisation to the datacube through a series of convolutions. Note that these convolutions are all applied to the time dimension, thus independently at each spatial pixel. Firstly, the datacube is convolved along the time dimension with a narrow Gaussian-shaped kernel of one-sigma width 24s, or 2 time steps, giving datacube  $D$. This initial convolution helps reduce noise. Secondly, $D$  is convolved along the time dimension with a wide Gaussian kernel $k$, of one-sigma width 150s or 12.5 time steps. This results in a time-smoothed datacube $D'$:
\begin{equation}
D' = D \otimes k,
\end{equation}
\noindent where the operation is applied to each spatial pixel's time series independently, with the convolution over time only. The choice of width for the narrow kernel is based on a qualitative visual inspection of the final images, with the width of the kernel kept at a minimum whilst maintaining coherent motions in the processed images. The choice of width for the wide kernel $k$ is investigated in detail in Section 5 of \cite{morgan2018}. They showed that the choice of width for $k$ has very little effect on the topology of the velocity field, although the velocity magnitude was sensitive to the choice of width.

A sliding window standard deviation of the signal over time at a given pixel, $\sigma$, is calculated as
\begin{equation}
\sigma = \sqrt{(D - D')^2 \otimes k}.
\end{equation}
\noindent Finally, the time-normalised series $D''$ is given by
\begin{equation}
D'' = \frac{D - D'}{\sigma + c}, 
\end{equation}
where $c$ is a small constant to reduce the effect of dividing by small numbers in very low-signal regions. In this work, $c=0.3$ \dns. For regions of study on the solar disk, this small value of $c$ has negligible effect since $\sigma \gg c$. It was included specifically to satisfy test cases where we applied the method to off-limb regions, where the signal becomes very low and $\sigma$ is small. Following this process, we discard an appropriate number of images (13 images) from the start and end of the series in order to avoid the edge effects of the convolutions.

Following time normalisation, each image in $D''$ is decomposed using a standard \atrous\ algorithm applied in the two spatial dimensions, then recomposed after discarding the finest scale composition \citep{starck1998}. This step serves to effectively discard some of the pixel-to-pixel noise, and leads to cleaner results. Figure \ref{fig2}b shows an example image, with the accompanying movie showing the time series. The time-normalised movie shows a myriad of small-scale motions at all times, and in all areas of the region. Whilst difficult to decipher directly by eye, the impression is gained of coherent regions of motion, and of a repetitive pattern.

\subsection{Modified Lucas-Kanade algorithm}
The final step in the process is to apply a modified Lucas-Kanade optical flow method \citep{lucas1981}, to characterise the motions in the time-normalised datacube. \citet{morgan2018} described a complicated approach where the velocity field was modelled using a truncated sine series for a local region in the image. The image was tiled into small square areas, the velocity fitted at each tile, then the tiles combined to give a velocity field over the whole region. This was applied incrementally to consecutive image pairs throughout the time series, then the final velocity field calculated through a weighted average over time (over all consecutive image pairs). This was a computationally cumbersome process, and the fitting of the velocity to a local function led to some interpretive uncertainty. That is, there was some uncertainty in the degree of smoothing in the resulting velocity fields. The following method replaces these steps with a simpler and far more efficient approach, and assumes that there is a dominant direction to the motion at a given pixel over time. This assumption is discussed in the context of the fitting residuals in section \ref{validity}.

The Lucas-Kanade method estimates the motion of objects in image pairs by assuming that the temporal change in image brightness is due solely to the motion. Thus the $x$ and $y$ derivatives of image brightness are related to the time derivative by:
\begin{equation}
\frac{\Delta I}{\Delta x}v_x + \frac{\Delta I}{\Delta y}v_y = -\frac{\Delta I}{\Delta t},
\label{eqnlk1}
\end{equation}
\noindent
where $v_x$ and $v_y$ are the $x$ and $y$ components of velocity, and $I$ is the image brightness. This equation cannot be solved for a single pixel (two unknowns), so typically the spatial and temporal derivatives of brightness are calculated for a local group of pixels, and the velocities estimated through the least-squares solution. In our method, we take advantage of the long time series (600 time steps), which gives a spatially local set of derivatives for a given pixel. 

The space and time derivatives of Equation \ref{eqnlk1} are all calculated by convolution of the datacube with a narrow kernel, independently across each dimension of the cube. The kernel, $k_d$, is the finite numerical derivative of a Gaussian of one-sigma width 1.5 (spatial pixel or time step), thus 
\begin{equation}
k_d=[0.01, 0.03, 0.10, 0.18, 0.16, 0.00, -0.16, -0.18, -0.10, -0.03, -0.01].
\end{equation}
\noindent The application of a Gaussian derivative $k_d$, which has a width spanning more than two pixels, provides a smoothing of the datacube along all dimensions. The issue of smoothing and the locality of solutions is inherent to optical flow methods, and is discussed in section 5 of \cite{morgan2018}. $k_d$ provides only a moderate smoothing essentially over $\pm3$ bins from the center (since the outer 2 values of $k_d$ are very small), thus the spatial velocity solution is influenced by a local group of approximately $7 \times 7$ pixels, which helps to alleviate the influence of noise, whilst giving sufficient resolution to the result. The convolution is applied over the $x$ dimension in order to gain $\Delta I / \Delta x$, and similarly for the $y$ and $t$ dimensions. Once the derivatives are found (resulting in three derivative cubes of the same size as the input datacube), the procedure applies least squares to solve Equation \ref{eqnlk1} individually for each pixel, giving solutions $v_x$ and $v_y$ at each pixel. 

This procedure describes a simple and efficient approach to estimate the optical velocities. We apply the following iterative approach:
\begin{enumerate}
\item Initialise master velocity arrays $V_x$ and $V_y$ as zero at all pixels. 
\item Begin iteration at iteration counter $i=0$.
\item If $i>0$, then use the current estimate of $V_x$ and $V_y$ to modify the datacube over time, essentially removing our current estimate of velocity translational shifts locally. Thus, if at a given pixel we have a velocity estimate of 0.5 pixel per time step in the positive $x$ direction, interpolation is used to form a new time series for that pixel, by extracting values from the datacube over time corresponding to positions of 0.5 pixels per time step. The modified datacube, $D_c$, is calculated from the initial datacube at each iteration $i>0$. For the first iteration, $D_c$ is set equal to the initial datacube (no interpolation).
\item Apply differencing kernel $k_d$ to $D_c$ to calculate the $x$, $y$, and $t$ derviatives.
\item Use least squares to solve for $v_x$ and $v_y$ in Equation \ref{eqnlk1}.
\item Use the current solution $v_x$ and $v_y$ to calculate the residual $\frac{\Delta I}{\Delta t}$ using Equation \ref{eqnlk1}.
\item Apply a weighted smoothing to $v_x$ and $v_y$. The velocity arrays are convolved with Gaussian kernels. This is done five times for kernels increasing in sigma width from 1 to 5 pixels, with the resulting smoothed velocity arrays recorded for each width, resulting in five arrays of increasing smoothness. Pixels with the smallest (largest) residuals will adopt a value from the least (most) smoothed array, with the magnitude of the residuals dictating the degree of smoothness.
\item Add the resulting velocities to the master velocity arrays $V_x$ and $V_y$. These are the current solutions.
\item Evaluate the velocity magnitude $V=\sqrt{V_x^2+V_y^2}$
\item Evaluate the mean absolute relative difference $m$ between the current estimate $V_i$ and the previous iteration's $V_{i-1}$.
\item If $m$ is greater than 20\%, and $i<6$, then continue iteration by repeating from step 3, else the process is terminated.
\end{enumerate}

Note that item 7 above smooths the solutions $v_x$ and $v_y$ spatially. This is a common approach in iterative optimisation \cite[e.g.][]{morgansites2019}, and a general guide is to apply as little smoothing as possible whilst avoiding the amplification of unphysical large magnitudes and discontinuities with iteration. Trial and error leads us to the approach of using the fitting residuals to determine the degree of smoothing, and the kernel widths of 1 to 5 pixels. For regions that best satisfy Equation \ref{eqnlk1}, the smoothing is minimal at 1 pixel width.

\subsection{Validity of the solution and uncertainties}
\label{validity}
Equation \ref{eqnlk1} is based on the assumption that the change in image brightness at a given pixel is due solely to motion. This is an inherent problem of optical flow methods. Whilst a visual inspection of the time-normalised movie associated with figure \ref{fig2}b clearly suggests coherent and repetitive motions, the changes in brightness can also be caused by other factors. Brightness can change in response to plasma density and temperature not associated with bulk plasma motions or waves, although the time normalisation step is a filter which effectively damps changes over longer than 150s. Brightness also changes randomly due to Poisson noise in the measurement, although the smoothing by a narrow kernel at the method outset helps to damp this high frequency component. Large-scale motions of coronal regions by differential rotation or sunspot rotation can also lead to brightness changes since our image alignment approach is purely translational, with a constant $x$ and $y$ pixel shift applied to the whole image. Any such large-scale motions will be present in the velocity results: section 5 and figure 15 of \cite{morgan2018} shows a clear statistical bias in the $x$ velocities related to solar rotation. Note that this new method includes the alignment step which effectively removes a constant solar rotation from the time series. This was absent from the previous method.

\begin{figure}    
\centerline{\includegraphics[width=1.0\textwidth,clip=]{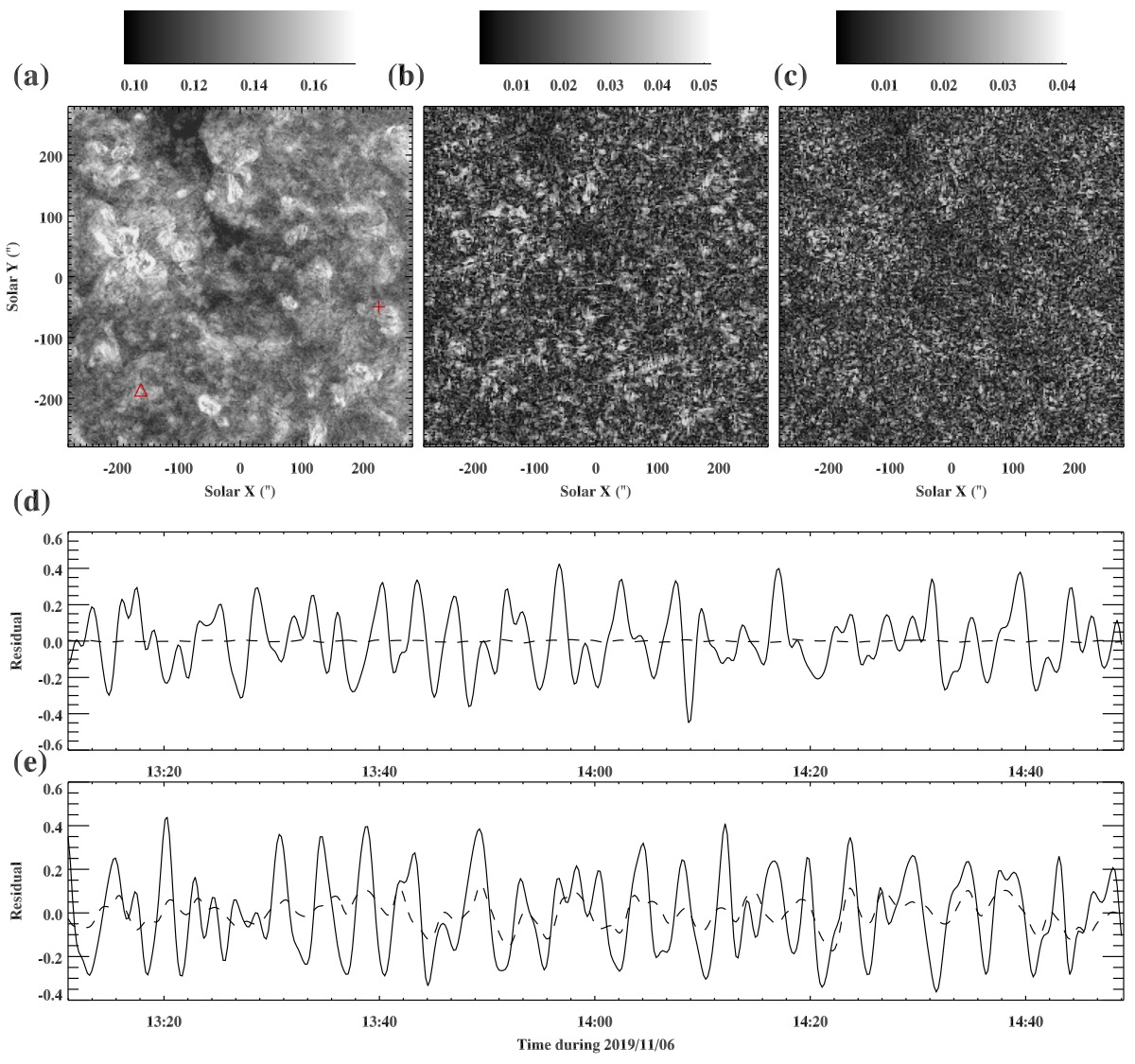}}
\caption{(a) The spatial distribution of the mean absolute residual (MAR) of Equation \ref{eqnlk1} when both $v_x$ and $v_y$ are set to zero (mean over all time), with the values indicated by the colour bar above the image. (b) The MAR after the first iteration. (c) The MAR after six iterations. (d) The residual over time for the pixel with the smallest final MAR. The initial residual is shown as a solid line, and the final as a dashed line. The position of this pixel is shown as a red cross in panel (a). (e) The initial (solid line) and final (dashed line) residual over time for the pixel with the largest final MAR. The position of this pixel is shown as a red triangle in panel (a).}
\label{residual}
\end{figure}

Under these limitations, Equation \ref{eqnlk1} is solved using the method described above, giving the $V_x$ and $V_y$ (smoothed) velocity field which best satisfies the equation in a least-squares sense. Figure \ref{residual} displays information on the residuals of Equation \ref{eqnlk1} at different steps of the iteration. Figure \ref{residual}a shows the spatial distribution of the initial mean absolute residual (MAR, or the mean over time of the absolute residuals). Figure \ref{residual}a thus gives an indication of the magnitude of the residual for a velocity field which is zero everywhere. Figures \ref{residual}b and c show the MAR after the initial and final (sixth) iteration respectively. Note that the least-squares smoothed velocity field greatly reduces the mean residuals after one iteration - thus a velocity field is rapidly found which satisfies Equation \ref{eqnlk1} better than a zero velocity field, and this is true across the whole image. Further iterations gradually reduce the residuals until a final limit is reached, that is, further iterations do not reduce the MAR. We can assume that the final residuals at a given pixel are therefore not caused by coherent motions in a single dominant direction, and must be caused by other temporal changes not associated with motions, and/or by occasional motions in different directions. 

\begin{figure}    
\centerline{\includegraphics[width=1.0\textwidth,clip=]{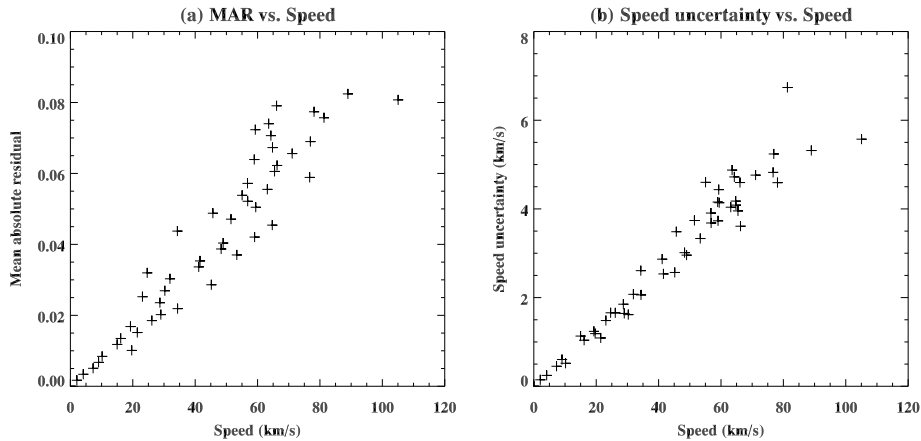}}
\caption{(a) The MAR as a function of speed for a selection of 50 pixels. (b) An estimate of the uncertainties in speed as a function of the speed for the same 50 pixels. This estimate is based on a bootstrapping approach based on randomisation of residuals following an initial least squares fitting to equation \ref{eqnlk1} (see text).}
\label{errors}
\end{figure}

Figures \ref{residual}d and e show the residual over time for the pixel with the smallest and largest final MAR respectively. The solid line shows the initial residual (zero velocity), and the dashed shows the residual at the final iteration. The velocity found for the pixel of figure \ref{residual}d obviously satisfies the changes in brightness well over all time, since the final residual is close to zero. The solution is less satisfactory for the pixel of figure \ref{residual}e. There is a clear linear relationship between the magnitude of the velocity at a given pixel and the MAR which is shown in figure \ref{errors}a. This figure shows, for a subset of 50 pixels, the MAR as a function of the speed at that pixel. 

Inspection of the distribution of the MAR can show the effectiveness of the velocity field in describing the changes in brightness. The distribution of MAR for the final velocity solution, shown in figure \ref{residual}c, varies greatly from pixel to pixel and is not smooth or coherent over large regions. This means that there are no large regions where the method is performing worse than for other regions. One exception to this is application of the method to large and complex active regions. In tests on active region data, we have noticed discrepancies in the velocity field where a coronal loop system crosses an underlying loop system. This is to be expected, since Equation \ref{eqnlk1} tries to find a single velocity which is smoothly consistent with the local group of pixels, whereas a superimposed system of loops at different orientations suggests two or more velocities are needed (at two or more heights in the corona).

An estimate of errors in the method is difficult. In particular, error propagation from the original data, through the time normalisation process, and our implementation of the Lucas Kanade algorithm is not feasible. A bootstrapping approach to error estimation is required, but is not feasible to apply to all pixels since the bootstrapping estimate requires repeating the least-squares fit hundreds of times to gain meaningful statistics. We instead select a subset of pixels covering the full range of MAR, and apply the randomised-residual bootstrapping approach described by \cite{byrne2013}. For each pixel, we find a least-squares solution to Equation \ref{eqnlk1} and calculate the residuals over time (as shown in figures \ref{residual}d and e). These residuals are randomised in order, and added to the original $\frac{\Delta I}{\Delta t}$ of Equation |ref{eqnlk1}. A least-squares solution is found, and the current $V_x$ and $V_y$ solutions recorded. This process is repeated 300 times, giving a distribution of solutions, with the spread based on the magnitude of the residuals. The standard deviation of the velocity distribution gives an error estimate on the resulting speed. Figure \ref{errors}b shows the speed uncertainty as a function of speed for a selection of 50 pixels. There is a clear linear relationship equating to a mean uncertainty of 6.5\%\ on our speed estimates. This is just an indication of the uncertainties as dictated by the fitting residuals, and does not include biases introduced by choices of various method parameters such as the choice of wide kernel width in the time normalisation. 

\subsection{Computational efficiency}
Using a standard desktop computer (3.7 GHz 6-Core Intel Core i5, with 32Gb memory), the whole process including reading in the data files, preprocessing, time normalisation, and the Lucas-Kanade algorithm, takes approximately 5 minutes. This is a factor of around 40 times faster than the implementation of \cite{morgan2018}, and enables rapid processing of multiple datasets for future studies. The Lucas-Kanade iterative process converges rapidly during the first two or three iterations, and terminates when the iteration counter reaches the set maximum number of iterations ($i=5$). \st{$m$ typically converges to a value of \app25\%. This value is likely related to the space and time derivatives being increasingly dominated by noise with increasing iteration.}

\section{Results}
\label{results}  

\begin{figure}    
\centerline{\includegraphics[width=0.9\textwidth,clip=]{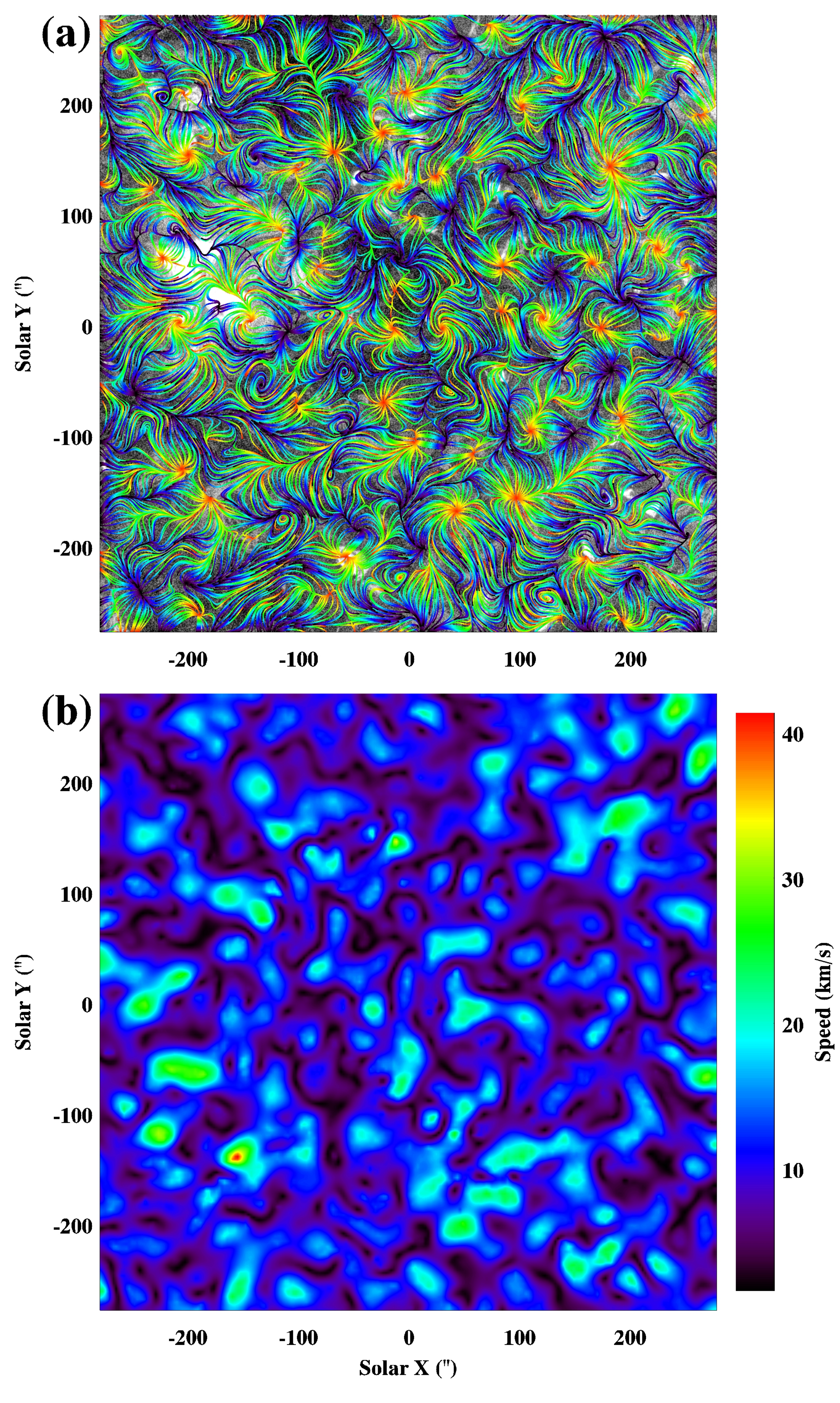}}
\caption{(a) The vector velocity field for the 193\AA\ channel across the region of interest. Each line is formed by starting from a random position, and tracing the field both backwards and forwards until a set number of steps is reached, or the length of the step becomes very close to zero (i.e. in regions of close to zero speed). The changing colour of a line along its length indicates the direction of flow, starting from red, advancing from yellow to green, and ending on blue. This field is superimposed on the background intensity image. (b) The speed in units of \kms, as indicated by the colour bar.}
\label{fig3}
\end{figure}

Applying the method to the 2019/11/06 193\AA\ channel dataset results in the velocity field shown in figure \ref{fig3}a. As shown by \cite{morgan2018}, the method can be applied to all the higher signal channels of AIA (304, 171, 193, and 211\AA), with differing results. For the sake of demonstrating the method, the 193\AA\ channel is chosen here since it has a high signal, and is dominated by emission from coronal plasma thus the results can be interpreted firmly in terms of coronal velocity fields, with little `contamination' from lower atmospheric layers. The most common pattern of the field distribution for this quiet Sun region is a network of distinct cells, approximately circular, and of typical diameter 50 to 100\arcsec\ (36 to 72Mm). The largest coherent cell is centered near $x,y=200,140$\arcsec, and has a diameter well over 100\arcsec. Flows tend to begin at the centers of the cells (red colours) and diverge towards the cell boundaries (blue). Another common pattern consists of long spines of converging or diverging flows. One example of a divergent spine is located near $x,y=150,-150$\arcsec. Some cells show rotation of flow, such as the cell located at $x,y=30,0$\arcsec, or another small cell at $x,y=-140,-60$\arcsec . 

Figure \ref{fig3}b shows the estimated speed given by the Lucas-Kanade algorithm. Maximum speeds are 40\kms, with most areas with speeds at 20\kms or less. The regions of brightest intensity in the 193\AA\ channel tend to correspond to the higher speeds, and also to regions of flow divergence (cell centers), although this relationship is not consistent across the whole region. The boundaries between cells are narrow and distinct. That is, there are no broad regions of close to zero speed, and cells bound each other closely. We do not have information on the line-of-sight motion, and the method gives only the plane of sky velocities. Furthermore, the results must be dominated by the atmospheric layer, or the height formation of the observed emission. The 193\AA\ channel has peak response near a temperature of 1.6MK, so is sensitive to the motions in the low corona. 

\subsection{Comparison to the photosphere}
\begin{figure}    
\centerline{\includegraphics[width=1.0\textwidth,clip=]{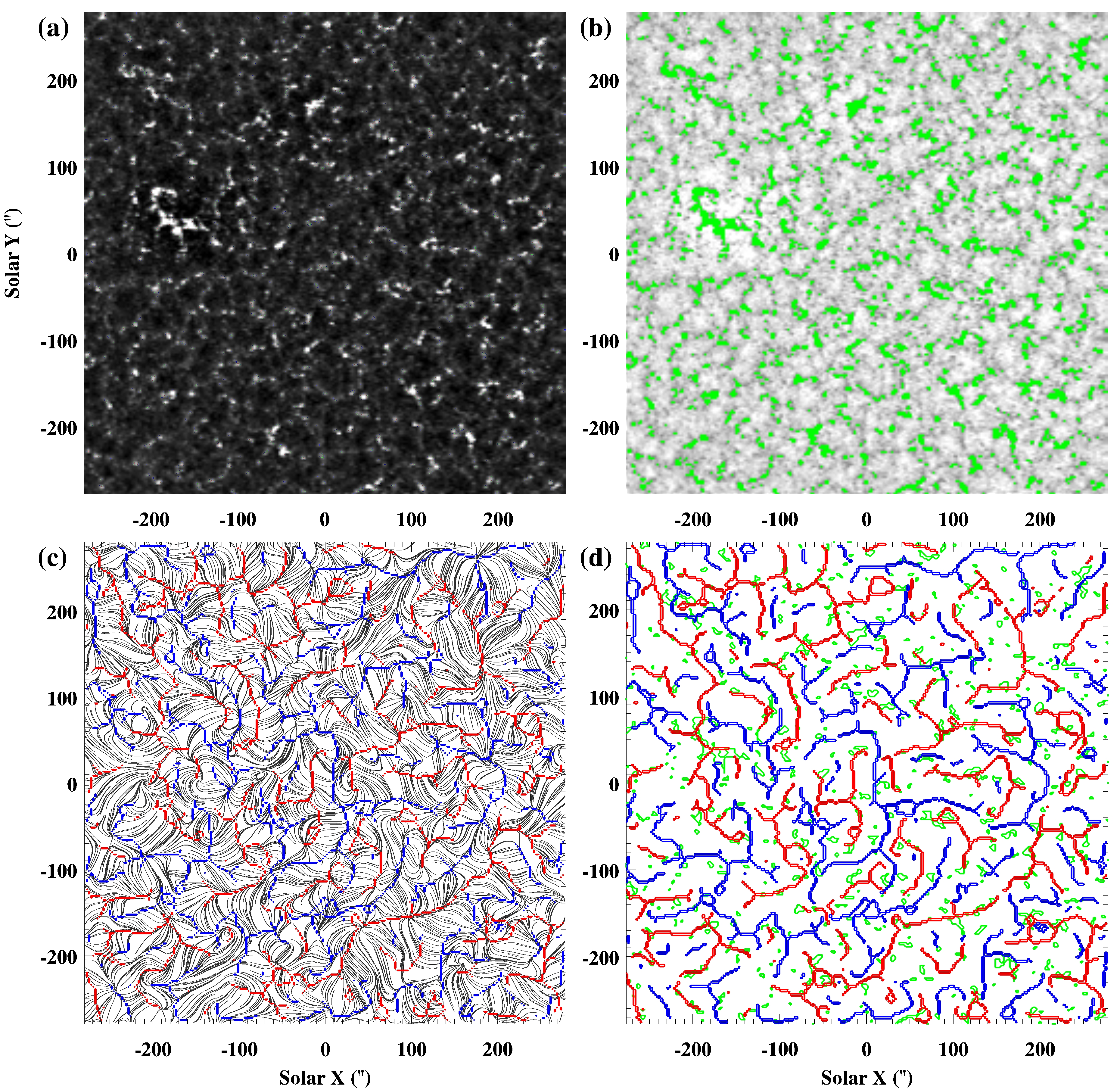}}
\caption{(a) Observation of the region of study by the AIA 1700\AA\ channel, showing the photospheric UV continuum. (b) The negative of the AIA 1700\AA\ channel intensity (white = low intensity), with the brightest network features highlighted in green. (c) Velocity vector map (black lines) shown without color or background image. The red and blue lines show where there is a large change in direction of the local velocity vector, with red corresponding to positive divergence (or source of flows) and blue corresponding to negative divergence (sinks, or where flows converge). (d) A comparison of the brightest UV network regions (green) with the AIA 193\AA\ velocity field characterisation of positive divergence (red) and negative divergence (blue). }
\label{fig4}
\end{figure}

Figure \ref{fig4}a shows an AIA 1700\AA\ image of the photospheric ultraviolet (UV) continuum. To create this image, we take 13 images taken every 10 minutes between 2019/11/06 13:00 to 15:00, apply image alignment so that every image is aligned to the 2019/11/06 14:00 image, then calculate an average image. Whilst the photosphere experiences rapid changes on small scales over a two hour period, the main bright network features, on spatial scales of tens of arcsecs, remain fairly consistent and stable over this period. Figure \ref{fig4}b shows the brightest features highlighted in green. The spatial scale of this distribution is consistent with the boundaries of the supergranular network.

We wish to compare this photospheric network with the topology of the coronal velocity field. In order to do this, we calculate a simplified representation of the complicated flow fields. Figure \ref{fig4}c shows the vector velocity field in black, and the blue and red lines represent certain regions where the direction of the flow field changes direction abruptly. The direction of the flow field at each pixel is given by the angle $\Omega = \arctan (v_y/v_x)$. The Sobel edge enhancement operator is applied to this angular image to give a value at each pixel: very high values are found at regions where the angle in velocity direction changes abruptly - both the red and blue lines in figure \ref{fig4}c correspond to high values. The red (blue) colours denote where these regions of abrupt angular change have positive (negative) divergence. The blue lines overlie regions where flows converge from surrounding areas (sinks), whilst the red lines overlie the points and corridors where flows diverge to surrounding areas (sources). 

Figure \ref{fig4}d plots the bright photospheric UV network in green, with the simplified representation of the coronal velocity field in red and blue. From a visual comparison, it is obvious that there is correlation between the distribution of photospheric network and the coronal velocity field. In many places, the position of photospheric and coronal features agree closely, for example at $X, Y = -200 , 50$\arcsec. There is obviously much further depth that can be applied to this analysis - for example, a rigorous correlation analysis, and a comparison with other AIA channels. Whilst this paper is more concerned with presenting the method and preliminary results, this qualitative comparison suggests that the coronal velocity field is strongly linked to the magnetic field topology in the quiet Sun. In particular, it seems as if the red lines, or the coronal sources, are most strongly correlated with the bright photospheric network. These preliminary results are closely linked to the findings of \cite{sheeley2012}, who found regions of clear cellular structures in AIA 193\AA\ images, linked to the underlying distribution of supergranular boundaries. 

\subsection{Comparison to the previous method}

\begin{figure}    
\centerline{\includegraphics[width=1.0\textwidth,clip=]{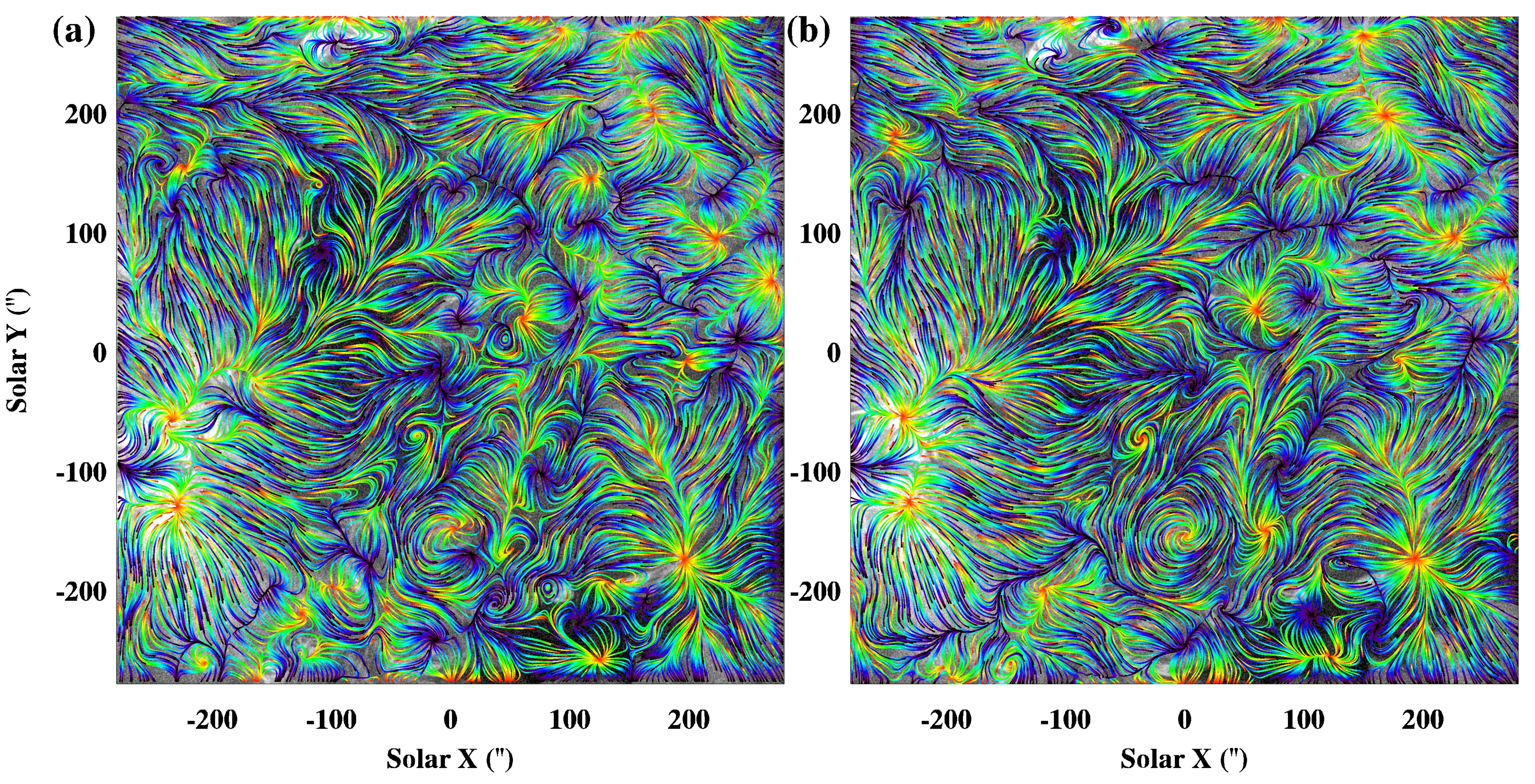}}
\caption{Comparison of the vector velocity fields for (a) this paper's method, and (b) the method of \cite{morgan2018}. These results are for a disk centre region observed by the AIA 193\AA\ channel during 2015/03/21 18:00 to 20:00, corresponding to the data used in \cite{morgan2018}.}
\label{fig5}
\end{figure}

Figure \ref{fig5}a shows the result of applying the new method to a disk centre region observed by the AIA 193\AA\ channel between 2015/03/21 18:00 to 20:00, corresponding to the data used in \cite{morgan2018}. This region has more varied structure compared to the quiet coronal example shown in figure \ref{fig3}, with a small active region in the south-east, bounded to the west by a dark region, possibly a dark canopy region \citep{wangrobbrecht2011} or a small coronal hole. Figure \ref{fig5}b shows the result of appyling the \cite{morgan2018} method to the same data. The two results are very similar across most of the region, with only small local differences in vector alignments and position. The new method shows a little more detail on smaller spatial scales, due to the old method's approach to model the velocities as smooth local geometrical functions.

\begin{figure}    
\centerline{\includegraphics[width=1.0\textwidth,clip=]{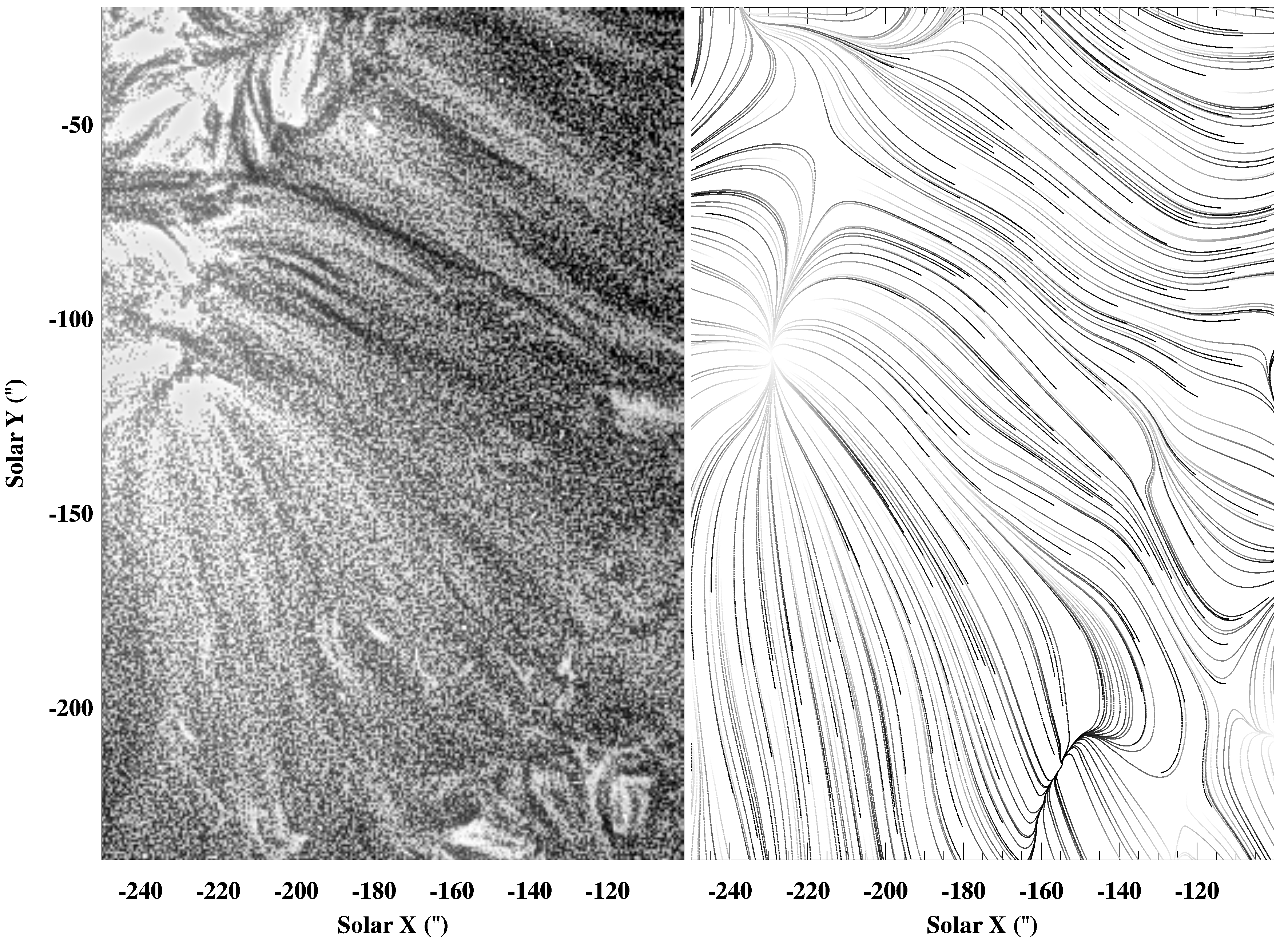}}
\caption{Left: An AIA 193\AA\ image of the small active region seen in the south east of figure \ref{fig5}, for an observation made at 2015/03/21 19:00. This image has been processed with multiscale Gaussian normalisation in order to enhance smaller-scale structures. Right: The vector velocity field calculated using the new method.}
\label{activeregion}
\end{figure}

The left panel of figure \ref{activeregion} shows more detail of the small active region in the south-east in the AIA 193\AA\ channel. This image has been processed using the multiscale Gaussian normalisation process to enhance smaller-scale features such as active region loops \citep{morgan2014}. The right panel shows the vector velocity field using the new method. Qualitatively, the agreement between the loops and structures extending outwards from the active region in the left image agrees well with the distribution of the velocity vectors in the right panel, which gives confidence that the method is indeed detecting and characterising coherent motions along magnetic structure. This type of direct comparison is difficult to make in the quiet Sun since the structure in the images is less clear, that is, it is difficult to see clear systems of large and distinct loops in the quiet Sun compared to active regions. 

\section{Summary}
\label{summary}

A time normalisation process applied to a series of AIA observations near disk centre reveals continuous and ubiquitous faint motions. A new, efficient method for tracing these motions is described. Application to quiet Sun observations of 2019/11/06 in the AIA 193\AA\ channel, dominated by hot coronal emission, gives speeds of 0-40\kms, and a velocity map which reveals a network of coherent flow cells of typical diameters 50 to 100\arcsec\ (36 to 72Mm). Results are similar to a previous, more complicated, method.

If we assume that the PD must follow the coronal magnetic field, then tracing their motions must give the direction of the magnetic field. We believe that the vector velocity field of figure \ref{fig3}a may give an indication of the coronal magnetic structure, and gives support to previous studies showing that the magnetic field of the quiet corona has a similar cellular structure. This is supported also by the visual comparison between the coronal velocity field and the photospheric UV network of figure \ref{fig4}d, which from a visual comparison seem to be strongly correlated. A comparison of the characteristics of the velocity field from multiple AIA channels with the photospheric network, to related photospheric motions, and photospheric vector magnetograms, is an approach which we are adopting to develop the results in future work. Another potential avenue is to combine the velocity vector fields with Doppler information from spectrometers, thus gaining a simultaneous map of plane-of-sky and line-of-sight speeds. We also plan to compare velocity fields for several AIA channels at different formation heights in order to gain further insight into the quiet Sun magnetic topology.

The software for this method is written in the Interactive Data Language (IDL), and can be used on any suitable datacube of regular cadence. The software is released as a package for public use at \url{https://github.com/HuwMorgan/TNOF}.

\begin{acks}
We acknowledge STFC grant ST/S000518/1 to Aberystwyth University, and the excellent facilities and support of SuperComputing Wales. The AIA/SDO data used here is courtesy of NASA/SDO and the AIA science team. 
\end{acks}
  
\bibliographystyle{spr-mp-sola}

\end{article} 

\end{document}